\documentclass[aps,showpacs,pra,twocolumn] {revtex4}

\usepackage{amsmath }
\usepackage{amssymb}
\usepackage{epsfig}
\usepackage{bm}

\input colordvi

\def\be{\begin{equation}}
\def\ee{\end{equation}}
\def\bea{\begin{eqnarray}}
\def\eea{\end{eqnarray}}

\def\ptl{\partial}

\begin{document}
\bibliographystyle{revtex}

\title{ Quantization of the inhomogeneous
Bianchi I model: quasi-Heisenberg picture }

\author{S.L. Cherkas}
\email{cherkas@inp.bsu.by}
 \affiliation{Institute for Nuclear Problems, Bobruiskaya
11, Minsk 220050, BELARUS}
\author{V.L. Kalashnikov}
\email{v.kalashnikov@tuwien.ac.at}
 \affiliation{Photonics Institute, Vienna University of Technology,
Gusshausstrasse 27/387, Vienna A-1040, AUSTRIA}

\received{ \today }

\begin{abstract}
The quantization scheme is suggested for a spatially inhomogeneous
1+1 Bianchi I model. The scheme consists in quantization of the
equations of motion and gives the operator (so-called
quasi-Heisenberg) equations describing an explicit evolution of a
system. Some particular gauge suitable for quantization is
proposed. The Wheeler-DeWitt equation is considered in the
vicinity of zero scale factor and it is used to construct a space,
where the quasi-Heisenberg operators act.
 Spatial discretization as a UV regularization
procedure is suggested for the equations of motion.
\end{abstract}

%Uncomment for PACS numbers title message
 \maketitle

\section{Introduction}

Spatially homogeneous minisuperspace  models \cite{hr,kie,rov} are
often used as a testbed for the quantum gravity
\cite{wheel,witt,CP,w,shest,hal}. Inhomogeneous 1+1 Bianchi I
model belongs to the so-called midisuperspace models \cite{bar}
and has more rich properties, in particular, it admits an
existence of gravitational waves. This model can be reduced  to
the Gowdy one \cite{gowdy} for which the solution of the
Wheeler-DeWitt equation has been obtained in a closed form
\cite{mizng,berger}. However, the solution of the Wheeler-DeWitt
equation does not resolve the problem of the gravity quantization
completely. An interpretation of the Wheeler-DeWitt equation
encounters the absence of a variable, which would play the role of
time, and all approaches to the quantum gravity face, as a rule,
this challenge.
 Some approaches regarding
the relation of the Wheeler-DeWitt equation to dynamics have been
suggested {\cite{is,hal1}}. Also, there exists a more
straightforward approach consisting in quantization of the
equations of motion \cite{prep1,prep2,gen}. The result of
quantization is the so-called quasi-Heisenberg operators.

Below we apply this approach to the quantization of the Bianchi I
model. The reason why
   we
 investigate the Bianchi I model instead of the Gowdy one
is that the former has a Hamiltonian, which is diagonal on the
momentums. Besides, it divides naturally a spatial geometry into
the scale factor and the remaining conformal geometry \cite{York},
while the Gowdy model suggests another separation.

Let us remind the quasi-Heisenberg picture in more details. It is
well-known that there are two equivalent quantum mechanical
pictures: Schr\"{o}dinger and Heisenberg ones. Both picture are
interrelated. In quantum gravity the situation is more
complicated. There is the Wheeler-DeWitt equation, which does not
admit any evolution.
 However, it is possible to obtain
an analog of the Heisenberg picture, that is so-called
quasi-Heisenberg picture. In fact it results from a direct
quantization of the classical equations of motion.  To quantize
the equations of motion one should write the classical equations
of motion and assume that every quantity is an operator. Then, one
has to choose some ordering of the operators in the equations of
motion. Later on, it should define the operator initial conditions
for the quasi-Heisenberg operators at an initial moment of time.
Thus, the constructed equations of motion and operator initial
conditions define the quantum evolution of a system completely.
Besides, the Hilbert space,  where the quasi-Heisenberg operators
act has to be builded.

The plan of this paper is the following: in Section 2, we explain
the basic structures of a classical nonuniform Bianchi model
including constraints algebra and evolution of constraints.  In
Section 3, we put forward the basic principles of the
quasi-Heisenbeg quantization scheme.  In Section 4, the
discretization of the equations of motion and operator ordering
are discussed. In Section 5, we expose how the results obtained
can be applied to the problem of vacuum state formation and to the
problem of vacuum energy.

\section{Nonuniform Bianchi model }

Let us consider the metric given by the interval

\begin{eqnarray}
ds^2=e^{2\alpha}\bigl(d\eta^2-e^{-4 B}dx^2-e^{2
B+2\sqrt{3}\,V}dy^2~~~~~~~~~~\nonumber\\~~~~-e^{2
B-2\sqrt{3}\,V}dz^2\bigr),
\end{eqnarray}
where the functions   $\alpha, B, V$ depend on the conformal time
$\eta$ and the spatial coordinate $x$. The spatially homogeneous
metric of such a type has been considered in Ref. \cite{mizn}.

Substitution of this metric into the Einstein equations allows
obtaining a set of five independent equations. Two of them are the
Hamiltonian and the momentum constraints
\begin{widetext}
\bea
 \mathcal H=\frac{1}{2}e^{2\alpha}\left(-(\ptl_\eta\alpha)^{
2}+(\ptl_\eta B)^{2}+(\ptl_\eta V)^{ 2}\right) +e^{2\alpha+4
B}\biggl(\frac{1}{6}(\ptl_x \alpha)^2
+\frac{1}{3}\ptl_{xx}\alpha+\frac{7}{6}(\ptl_x
B)^2~~~~~~~~~~~~\nonumber\\+\frac{1}{3}\ptl_{xx}B+\frac{4}{3}\ptl_x\alpha\ptl_x
B+\frac{1}{2}(\ptl_x V)^2\biggr)=0,\label{hamcon}\\
   \mathcal P=
e^{2\alpha}\left(-\frac{1}{3}\ptl_x\alpha\ptl_\eta\alpha+\ptl_x B
\ptl_\eta B+\frac{2}{3}\ptl_x\alpha\ptl_\eta B+\ptl_x V \ptl_\eta
V+\frac{1}{3}\ptl_x\ptl_\eta
B+\frac{1}{3}\ptl_x\ptl_\eta\alpha\right)=0,~~~~~~~~
\label{pconcon}
\eea
\end{widetext} and the rest are the equations of motion
\begin{widetext}
\bea
   \ptl_{\eta\eta}\alpha-e^{4B}\left(\ptl_{xx}\alpha
 +(\ptl_{x}\alpha)^2+(\ptl_x V)^2+
\frac{7}{3}(\ptl_x
B)^2+\frac{2}{3}\ptl_{xx}B+4\ptl_xB\ptl_x\alpha\right)
+(\ptl_\eta\alpha)^{ 2}+ (\ptl_\eta V)^{ 2}+(\ptl_\eta B)^{2}=0,\label{11}\\
\ptl_{\eta\eta}B+\frac{1}{3}e^{4B}\left(\ptl_{xx}B+2(\ptl_{x}B)^2+6(\ptl_x
V)^2-2(\ptl_x\alpha)^2+2\ptl_x\alpha\ptl_xB+2\ptl_{xx}\alpha\right)
+2
\ptl_\eta B\ptl_\eta\alpha=0,\label{12}\\
  \ptl_{\eta\eta}V+2 \ptl_\eta
V\ptl_\eta\alpha-e^{4B}\left(\ptl_{xx}
V+2\ptl_xV\ptl_x\alpha+4\ptl_xB\ptl_xV\right)=0.~~~~~~~~~~
~~~~~~~~~~~~~ \label{10}
\eea
\end{widetext}
\noindent The full Hamiltonian $H=\int \mathcal H dx$ has to be
zero during an evolution of system.

The relevant Hamiltonian and the momentum constraints written in
the terms of momentums $\pi_V\equiv\frac{\delta H}{\delta( {
\ptl_\eta V })}=e^{2\alpha}\ptl_\eta V$, $\pi_B\equiv\frac{\delta
H}{\delta {( \ptl_\eta B) }}=e^{2\alpha}\ptl_\eta B$
 and
$p_\alpha\equiv-\frac{\delta H}{\delta
(\ptl_\eta\alpha)}=e^{2\alpha}\ptl_\eta\alpha$ take the form

\begin{widetext}
\bea
  \mathcal H=\frac{1}{2}e^{-2\alpha}\left(-p_\alpha^{ 2}+\pi_B^{
2}+\pi_V^{ 2}\right) +e^{2\alpha+4 B}\left(\frac{1}{6}(\ptl_x
\alpha)^2+\frac{1}{3}\ptl_{xx}\alpha+\frac{7}{6}(\ptl_x
B)^2+\frac{1}{3}\ptl_{xx}B+\frac{4}{3}\ptl_x\alpha\ptl_x
B+\frac{1}{2}(\ptl_x V)^2\right),\\
  \mathcal P=-p_\alpha\ptl_x\alpha+\pi_B\ptl_x B+\pi_V\ptl_x
V+\frac{1}{3}\ptl_x \pi_B+\frac{1}{3}\ptl_x
p_\alpha.~~~~~~~~~~~~~~~~~~~~~~~~~~~~~~~~~~~~~~~~~~~~~~~~~~~~~~~~~~~~~
\eea
\end{widetext}

 Using the Poisson brackets

\bea
  \{F(x),G(x^\prime)\}=\int\Biggl(-\frac{\delta
F(x)}{\delta p_\alpha(\xi)}\frac{\delta G(x^\prime)}{\delta
\alpha(\xi)}+\frac{\delta F(x)}{\delta \alpha(\xi)}\frac{\delta
G(x^\prime)}{\delta p_\alpha(\xi)}\nonumber\\ +\frac{\delta
F(x)}{\delta \pi_V(\xi)}\frac{\delta G(x^\prime)}{\delta
V(\xi)}-\frac{\delta F(x)}{\delta
V(\xi)}\frac{\delta G(x^\prime)}{\delta \pi_V(\xi)}~~~~~~~~\nonumber\\
+\frac{\delta F(x)}{\delta \pi_B(\xi)}\frac{\delta
G(x^\prime)}{\delta B(\xi)}-\frac{\delta F(x)}{\delta
B(\xi)}\frac{\delta G(x^\prime)}{\delta \pi_B(\xi)}
\Biggr)d\xi~~~~ \label{pois} \eea

\noindent one may obtain the constraint algebra:
\[
\{\mathcal H(x),\mathcal H(x^\prime)\}=(\mathcal P(x)e^{4
B(x)}+\mathcal P(x^\prime)e^{4
B(x^\prime)})\delta^\prime(x^\prime-x),
\]
\bea
\{\mathcal P(x),\mathcal P(x^\prime)\}=(\mathcal P(x)+\mathcal
P(x^\prime))\delta^\prime(x^\prime-x),\nonumber\\ \{\mathcal
H(x),\mathcal P(x^\prime)\}=\frac{2}{3}(\mathcal H(x)+\mathcal
H(x^\prime))\delta^\prime(x^\prime-x)\nonumber\\-\frac{\mathcal
H^\prime(x)}{3}\delta(x^\prime-x)\nonumber.
\end{eqnarray}

It is also possible to find the evolution of constraints by
calculation of their Poisson brackets with the Hamiltonian $H$:

\begin{eqnarray}
\ptl_\eta\mathcal P(\eta,x)=\{H,\mathcal
P(\eta,x)\}=\frac{1}{3}\ptl_x \mathcal H(\eta,x),\label{sv1}\\
\ptl_\eta\mathcal H(\eta,x)=\{H,\mathcal
H(\eta,x)\}=\ptl_x\left(e^{4B(\eta,x)}\mathcal
P(\eta,x)\right).\label{sv2}
\end{eqnarray}

\section{Quantization}

The quantization procedure consists of two stages. At the first
stage we formulate the initial conditions for the quasi-Heisenberg
operators using the Dirac brackets \cite{dirac}. Thereafter it
will be permissible for operators to evolve in accordance with the
equations of motion, considered as the operator equations. To
determine the initial commutation relations, we will use the Dirac
quantization procedure \cite{dirac,han,git} where besides the
constraints, an additional gauge condition is needed.

Let us take the following gauge at the initial moment of time
\be
\mathcal A= \alpha-\alpha_0=0, \label{condal}
\ee
\be
\mathcal B=\ptl_x \pi_B=0,\label{condB}
\ee
where $\alpha_0$ should be tended to $-\infty$ finally.

For quantization by means of the Dirac brackets \cite{dirac}, one
has to calculate the matrix
$M_{ij}(x,x^\prime)=\{\Phi_i(x),\Phi_j(x^\prime)\}$, where a set
of constraints is
\[
\Phi_{i}=(\mathcal H,\mathcal P, \mathcal A,\mathcal B).
\]

At the shell of constraints $\Phi_i(x)=0$,  the matrix
$M_{ij}(x,x^\prime)$ in the vicinity of
$\alpha_0\rightarrow-\infty$ has the form
\begin{widetext}
\bea
 \bm M(x,x^\prime)=\left(
\begin{array}{cccc}
0&0&\frac{p_\alpha(x)\delta(x-x^\prime)}{\exp({2\alpha_0})}&0\\
0&0&-\frac{\delta^\prime(x-x^\prime)}{3}&\pi_B\delta^{\prime\prime}(x-x^\prime)\\
-\frac{p_\alpha(x)\delta(x-x^\prime)}{\exp({2\alpha_0})}&-\frac{\delta^\prime(x-x^\prime)}{3}&0&0\\
0&-\pi_B\delta^{\prime\prime}(x-x^\prime)&0&0
\end{array}
\right).
\eea
\end{widetext}
\noindent The primed functions correspond to their differentiation
over an argument.

Due to antisymmetry of the Poison brackets, $M_{ij}(x,x^\prime)$
obeys the identity $M_{ij}(x,x^\prime)=-M_{j\,i}(x^\prime,x)$. The
inverse matrix satisfying $\int
M_{ij}(x,x^{\prime\prime})M_{j\,k}^{-1}
(x^{\prime\prime},x^{\prime})dx^{\prime\prime}=\delta_{ik}\delta(x-x^\prime)$
is given by
\begin{widetext}
\[
 \bm M^{-1}(x,x^\prime)=\left(
\begin{array}{cccc}
0&0&-e^{2\alpha_0}\frac{\delta(x-x^\prime)}{p_\alpha(x)}&e^{2\alpha_0}\frac{\theta(x-x^\prime)}{3p_\alpha(x)\pi_B}\\
0&0&0&-\frac{\Delta(x-x^\prime)}{\pi_B}\\
e^{2\alpha_0}\frac{\delta(x-x^\prime)}{p_\alpha(x)}&0&0&0\\
-e^{2\alpha_0}\frac{\theta(x^\prime-x)}{3p_\alpha(x^\prime)\pi_B}&\frac{\Delta(x^\prime-x)}{\pi_B}&0&0
\end{array}
\right),
\]
\end{widetext}
where $\Delta(x-x^\prime)=|x-x^\prime|/2$ is a Green function of
the one dimensional Laplace operator:
$\frac{d^2\Delta}{dx^2}=\delta(x-x^\prime)$ and
$\theta(x-x^\prime)=\Delta^\prime(x-x^\prime)$.

According to  Ref. \cite{dirac}, the Dirac brackets result in the
commutation relations for the corresponding operators at an
initial moment of time after multiplication by $-i$.

Calculation of the Dirac brackets

\begin{widetext}
\begin{eqnarray*}
 \{G(x),F(x^\prime)\}_D=\{G(x),F(x^\prime)\}
-\sum_{i,j}\int \{G(x),\Phi_i(x^{\prime\prime})\}
M^{-1}_{ij}(x^{\prime\prime},x^{\prime\prime\prime})\{\Phi_j(x^{\prime\prime\prime}),F(x^\prime)\}dx^{\prime\prime}dx^{\prime\prime\prime}
\end{eqnarray*}
\end{widetext}

leads to

\be
\{\pi_V(x),V(x^\prime)\}_D=\delta(x-x^\prime),\nonumber
\ee
\bea
\{\pi_B(x),B(x^\prime)\}_D=0,~~~~~~~~~~~~~~~~~~~~~~~~~~~~~~~~~~\nonumber
\\
\{p_\alpha(x),\alpha(x^\prime)\}_D=0,~~~~~~~~~~~~~~~~~~~~~~~~~~~~~~~~~~~\nonumber\\
\{\pi_B(x),V(x^\prime)\}_D=0,~~~~~~~~~~~~~~~~~~~~~~~~~~~~~~~~~~\nonumber\\
\{\pi_B(x),p_\alpha(x^\prime)\}_D=0,~~~~~~~~~~~~~~~~~~~~~~~~~~~~~~~~~\nonumber\\
\{B(x),V(x^\prime)\}_D=~~~~~~~~~~~~~~~~~~~~~~
~~~~~~~~~~~~~~~~\nonumber\\-\frac{1}{\pi_B}\left(\theta(x-x^\prime)\ptl_{x^\prime}V(x^\prime)
+\frac{\pi_V(x)}{3p_\alpha(x)}\delta(x-x^\prime)\right),
\nonumber\\
\{B(x),p_\alpha(x^\prime)\}_D=~~~~~~~~~~~~~~~~~~~~~~~~~~~~~~~~~~~~~\nonumber
\\~~~~\frac{1}{\pi_B}\left(\ptl_{x^{\prime}}p_\alpha(x^\prime)\theta(x^\prime-x)
+\frac{\pi_V^2(x)}{p_\alpha(x)}\delta(x^\prime-x)\right),\nonumber\\
\{p_\alpha(x),V(x^\prime)\}_D=
\frac{\pi_V(x)}{p_\alpha(x)}\delta(x^\prime-x),~~~~~~~~~~~~~~~\nonumber\\
\{\pi_B(x),\alpha(x^\prime)\}_D=0.~~~~~~~~~~~~~~~~~~~~~~~~~~~~~~~~~~
\label{cm}
\eea

From the foregoing it follows that  $\alpha$ and  $\pi_B$ are
initially  $c$-numbers in the gauge considered. In fact, we use
some time-dependent gauge. Such a gauge is known only at an
initial moment of time. Hence, the commutation relations in the
model under consideration can evolve.

Operator realization of the commutation relations  at the initial
moment of time (and at $\alpha_0\rightarrow -\infty$) may be
written as
\[
\hat \pi_V(x)=-i\frac{\delta}{\delta V(x)},
\]

\[
\hat p_\alpha(x)=\sqrt{\hat \pi_V^2(x)+\pi_B^2},
\]

\bea
\hat B(x)=B_0-\frac{1}{\pi_B}\biggl(\int_{-\infty}^\infty
\theta(x-x^\prime)S(\hat
\pi_V(x^\prime)\ptl_{x^\prime}V(x^\prime))dx^\prime
\nonumber\\+\frac{1}{3}\hat p_\alpha(x)\biggr),\nonumber
\eea

\noindent where the symbol $S$ denotes symmetrization of the
noncommutative operators, i.e. $S(\hat A \hat B)=\frac{1}{2}(\hat
A\hat B+\hat B\hat A )$ or $S(\hat A\hat B\hat C)=\frac{1}{6}(\hat
A\hat B\hat C+\hat B\hat A\hat C+\hat A\hat C\hat B+\dots)$, and
$B_0$, $\pi_B$, $\alpha(x)=\alpha_0$ are some $c$-number
constants.

It is important to note that the spatially uniform quantities
$B_0$, $\pi_B$, $\alpha_0$ turn out to be the $c$-numbers, but not
the operators. Otherwise, the commutation relations (\ref{cm})
will have some constants besides $\delta$ and $\theta$ functions
on the right hand sides. The crucial point here is that an
infinite system is considered. Indeed, if a finite system like a
finite string is quantized, some constants appear already in the
matrix $M$ (see, for instance, Eq. (4.50) from Ref. \cite{han}).
It leads to appearance of the quantum spatially uniform variables,
which correspond to motion of the string center of mass. From this
standpoint, the approach of Ref. \cite{Ah} to a spatially uniform
quantized background seems to be appropriate for a closed
universe, which is finite by definition. Nevertheless, such an
approach is inadequate for an open and flat universe, which is
infinite.

Thus the equation of motion (\ref{11}),(\ref{12}),(\ref{10})
should be considered as the operator equations with the initial
conditions
\bea
 \hat V(0,x)=V(x),~~~~~~~~~~~~~~~~~~~~~~~
 ~~~~~~~~~~~~~~~~~~~\nonumber
 \\\hat V^\prime(0,x)=-i
\,e^{-2\alpha_0}\frac{\delta }{\delta
V(x)},~~~~~~~~~~~~~~~~~~~~~~~~~~~~
\nonumber\\
 \hat B(0,x)=B_0~~~~~~~~~~~~
 ~~~~~~~~~~~~~~~~~~~~~~~~~~~~~~~~~~
 \nonumber\\-\frac{1}{\pi_B}\biggl(\int_{-\infty}^\infty
\theta(x-x^\prime)S\biggl(-i \frac{\delta }{\delta
V(x^\prime)}\,\ptl_{x^\prime}V(x^\prime)\biggr)dx^\prime
\nonumber\\+\frac{1}{3}\sqrt{-\frac{\delta^2 }{\delta
V^2(x)}+\pi_B^2}\biggr),
\nonumber\\
 \hat B^\prime(0,x)=e^{-2\alpha_0}\pi_B,~~~~~~~~~~~~~~~~~~~~~~~~~~~~~~~~~~~~
 \nonumber\\\hat
\alpha(0,x)=\alpha_0,~~~~~~~~~~~~~~~~~~~~~~~~~~~~~~~~~~~~~~~~~~~~~
\nonumber\\
\hat \alpha^\prime(0,x)=e^{-2\alpha_0}\sqrt{-\frac{\delta^2
}{\delta V^2(x)}+\pi_B^2},~~~~~~~~~~~~~~~~~~ \label{rlz2}
\eea
where $\pi_B, B_0$ are some constants and $\alpha_0$ should be
tended to $-\infty$.

Rewriting Eqs. (\ref{rlz2}) in the momentum representation, where
$\hat \pi_V(x)=\pi_V(x)$ and $\hat V(x)=i\frac{\delta }{\delta
\pi_V(x)} $ gives

\bea
 \hat V(0,x)=i\frac{\delta }{\delta \pi_V(x)},~~~~\hat
V^\prime(0,x)= e^{-2\alpha_0}\pi_V(x),~~~~
\nonumber\\
 \hat B(0,x)=B_0~~~~~~~~~~~~~~~~~
 ~~~~~~~~~~~~
 ~~~~~~~~~~~~~~~~~~~
 \nonumber\\-\frac{1}{\pi_B}\biggl(\int_{-\infty}^\infty
\theta(x-x^\prime)S\biggl(\pi_V(x^\prime)\,i\ptl_{x^\prime}\frac{\delta
}{\delta \pi_V(x^\prime)}\biggr)dx^\prime
\nonumber\\+\frac{1}{3}\sqrt{\pi_V^2(x)+\pi_B^2}\biggr),
\nonumber\\
 \hat B^\prime(0,x)=e^{-2\alpha_0}\pi_B,~~~~\hat
\alpha(0,x)=\alpha_0,~~~~~~~~~~~~~~~~~\nonumber\\ \hat
\alpha^\prime(0,x)=e^{-2\alpha_0}\sqrt{\pi_V^2(x)+\pi_B^2}.~~~~~~~~~~~~~~~~~~~~~~~~
\label{rlz3}
\eea

The second stage of quantization procedure consists in building of
the Hilbert space, where the quasi-Heisenberg operators act. At
this stage we return to the classical Hamiltonian (\ref{hamcon})
and momentum (\ref{pconcon}) constraints. The momentum constraint
and the corresponding gauge condition (\ref{condB}) are resolved
relatively the variable $B$ and its momentum $\pi_B$. Then, these
quantities are substituted to the Hamiltonian constraint, which is
then quantized and considered as the Wheeler-DeWitt equation in
the vicinity of a small scale factor $a\sim 0$, i.e. $\ln
a=\alpha\rightarrow-\infty$. Thus, we come to
\begin{equation}
\left(\frac{\delta^2}{\delta \alpha(x)}-\frac{\delta^2}{\delta^2
V(x)}+\pi_B^2\right)\Psi[\alpha,V]=0, \label{witt}
\end{equation}
where it is taken into account that $\pi_B$ is some constant.
Space of the negative frequency solutions of (\ref{witt}) forms
the Hilbert space for the quasi-Heisenberg operators.

In the general case, a solution of Eq. (\ref{witt}) has the form
of the wave packet
\begin{eqnarray}
\Psi[\alpha,V]=~~~~~~~~~~~~~~~~~~~~~~ ~~~~~~~~~~~~~~~~ ~~~~~~~~
~~~~~~~~~~~\\\int C[\pi_V]\,e^{\int\left(-i
\alpha(x)\sqrt{\pi_B^2+\pi_V^2(x) }+i\pi_V(x) V(x)\right)d
x}\,\mathcal D \pi_V,\nonumber
\end{eqnarray}
where $\mathcal D \pi_V$ denotes the functional integration over
$\pi_V(x)$ and only negative frequency solutions are taken.

Scalar product has a form
  \cite{witt,prep1, prep2, gen}
\begin{widetext}
\begin{eqnarray}
   <\Psi|\Psi>=i\, Z \prod_x\int \biggl( \Psi^*[\alpha,V] {\hat
D^{-1/2}}(x)\frac{\delta}{\delta \alpha(x)
}\Psi[\alpha,V]-\left({\hat D^{-1/2}}(x)\frac{\delta}{\delta
\alpha(x) }\Psi^*[\alpha,V]\right)\Psi[\alpha,V]\biggr)d
V(x)\biggr|_{\,\alpha(x)=\alpha_0\rightarrow -\infty }
\label{scal}
\end{eqnarray}
\end{widetext}
\noindent where $\hat D(x)=-\frac{\delta^2}{\delta
V^2(x)}+\pi_B^2$ and $Z$ is a normalization constant. Here, the
infinite product is taken over $x$ and has to be understood as a
result of limiting process on a $x$-space lattice.

Mean value of an arbitrary operator can be evaluated as
\begin{widetext}
\begin{eqnarray}
   <\Psi|\hat A[\alpha,-i\frac{\delta}{\delta V},V]|\Psi>=i \,Z
\prod_x\int \biggl( \Psi^*[\alpha,V] \hat A\,{\hat
D^{-1/2}}(x)\frac{\delta}{\delta \alpha(x)
}\Psi[\alpha,V]~~~~~~~~~~~~~~~~~~~~~~~~~~~~~~~~~~~~~~~~~~~~~~~~~~
\label{mean1}\\-\left({\hat D^{-1/2}}(x)\frac{\delta}{\delta
\alpha(x) }\Psi^*[\alpha,V]\right)\hat A\,\Psi[\alpha,V]\biggr)d
V(x)\biggr|_{\,\alpha(x)=\alpha_0\rightarrow -\infty }.\nonumber
\end{eqnarray}
\end{widetext}

Let us note that the hyperplane $\alpha(x)=\alpha_0$ along which
the integration is performed in (\ref{mean1}) coincides with the
initial condition for the quasi-Heisenberg operator $\hat \alpha$
in (17). The most convenient momentum representation  results in
the wave function $\psi$
\begin{equation}
{\psi}[\alpha,\pi_V]=C[\pi_V]\exp\left({-i\int\alpha(x)\sqrt{\pi_V^2(x)+\pi_B^2}\,dx}\right).
\end{equation}
Then, a mean value of an operator becomes

\begin{eqnarray}
  <\psi|\hat A[\alpha,\pi_V(x),i\frac{\delta}{\delta
\pi_V(x)}|\psi>=~~~~~~~~~~~~~~~~~~~
~~~~~~~~\nonumber\\
 \label{mean2}\nonumber\\\int
C^*[\pi_V]e^{-i \int \alpha(x)\sqrt{\pi_V^2(x)+\pi_B^2} \,dx}\hat
A\,e^{i \int
\alpha(x)\sqrt{\pi_V^2(x)+\pi_B^2}\,dx}\nonumber\\C[\pi_V]\,\mathcal
D \pi_V\biggl|_{\,\alpha(x)=\alpha_0\rightarrow -\infty }.~~~~~
\end{eqnarray}

Thus, one has an exact quantization scheme consisting of the
Wheeler-DeWitt equation in the vicinity of small scale factor
(\ref{witt}), the operator initial conditions (\ref{rlz2}) for the
equations of motion (\ref{11},\ref{12},\ref{10}) and the
expressions (\ref{mean1}), (\ref{mean2}) for calculation of the
mean values of operators.

\section{Discretization of the operator equations}

Although the above quantization scheme is exact it can not be
applied for numerical calculations, because infinite quantities
will arise if one attempts to calculate some mean values by means
functional integration. At least two generations of physicists can
not overcome divergencies arising in rigorous operator formulation
of the ordinary QFT in the 4-dimensional Minkowsky space, although
some success for the (1+1) and (1+3) -- dimensional models has
been reached \cite{jaffe}. To overcome these difficulties, we
purpose a discretization method for regularization of the
functional operator equations.  The elementary discretization
consists in choosing of some spatial box of length $L$ and
granulation of it by points $x_i$ separated by distance $\Delta
x$. Periodicity condition is implied at $x_{N+1}=x_1$. In
principle it is more fundamental to write the discretizied action
initially and then quantize the system. But at this way one may
encounter the some additional complexity, because some additional
constraints could arise besides the Hamiltonian and momentum
constraints, as it happens for the quantization of the
discretizied  string \cite{tmf}. Additional constraints will not
be essential because they will vanish in the continuous limit and
only hamper an analysis. Thus, we prefer to do the discretization
empirically  looking at the exact (but practically unusable)
functional equations obtained in the previous section.

Continuous $V$-field and its momentum $\pi_V(x)$  should be
replaced by the discrete quantities $ \pi_V(x)\rightarrow
\pi_{Vj}/\sqrt{\Delta x}$, $V(x)\rightarrow V_j /\sqrt{\Delta x}$
\cite{tirr}, where $\pi_{Vj}$ and $V_j$ posses an ordinary
commutation relation $[\pi_{Vn},V_m]=-i\delta_{nm}$ including the
Kronecker symbol $\delta_{mn}$. Indeed, we have
$[\pi_{V}(x_n),V(x_m)]=-i\delta_{nm}/\Delta x$ in this case for
$\pi_V(x_n)$ and $V(x_m)$, that turns into the Dirac
delta-function $\delta_{nm}/\Delta x\rightarrow \delta(x_n-x_m)$
in the limit of $\Delta x\rightarrow 0$. However, it is more
convenient to use the quantities $\pi_{Vj}= \pi_V(x_j)$,
$V_j=V(x_j)$ straightforwardly, for which the commutation
relations $[\pi_{Vn},V_m]=-i\delta_{nm}/\Delta x$ should be used,
as well. These commutation relations can be realized by the
operators $\hat V_m=V_m $, $\hat\pi_{Vn}=-\frac{i}{\Delta x}
\frac{\ptl}{\ptl V_n}$ , or alternatively by $\hat
\pi_{Vm}=\pi_{Vm} $, $\hat V_n=\frac{i}{\Delta x} \frac{\ptl}{\ptl
\pi_{Vn}}$.

The discrete Wheeler-DeWitt equation in the vicinity of
$\alpha_j=\alpha_0\rightarrow - \infty$ has the following form in
the momentum representation:

\bea
\left(-\frac{1}{(\Delta x)^2}\frac{\ptl^2}{\ptl^2
\alpha_i}+\pi_{Vi}^2+\pi_B^2\right)\psi(\alpha_i,\dots\alpha_N,\pi_{V1},\dots
\pi_{VN})\nonumber\\=0, \nonumber
\eea

\noindent with the solution
\bea
  {\psi}(\alpha_1\dots\alpha_N,\pi_{V1},\dots,\pi_{VN})=~~~~~~~~~~~
  ~~~~~~~~~~~~~~~~~~~~~~~
\nonumber\\
C(\pi_{V1},\dots...,\pi_{VN})\exp\left({-i\Delta x
\sum_{j=1}^N\alpha_j\sqrt{\pi_{Vj}^2+\pi_B^2}}\right),~~
\eea
where $\Delta x$ is the discretization length.

A mean value of an arbitrary operator can be evaluated as
\begin{widetext}
\begin{eqnarray}
  <\psi|\hat A(\alpha_1\dots\alpha_N,\pi_{V1},
 \dots\pi_{VN},\frac{i}{\Delta x}\frac{\ptl}{\ptl \pi_{V1}},\dots ,\frac{i}{\Delta x}\frac{\ptl}{\ptl \pi_{VN}})|\psi>
 =~~~~~~~~~~~~~~~~~~~~~~~~~~~~~~~~~~~~~~~~~
 ~~~~~~~~~~~~~~~~~~~~~~~~~~
 \nonumber\\\int
C^*(\pi_{V1}\dots\pi_{VN})e^{-i\Delta x \sum_{j=1}^N
\alpha_j\sqrt{\pi_{Vj}^2+\pi_B^2} }\hat A\,e^{i \Delta x
\sum_{j=1}^N
\alpha_j\sqrt{\pi_{Vj}^2+\pi_B^2}}C(\pi_{V1}\dots\pi_{VN})d\pi_{V1}\dots
d\pi_{VN} \biggl|_{\,\alpha_1\dots\alpha_N=\alpha_0\rightarrow
-\infty }.
\end{eqnarray}
\end{widetext}

For the equations of motion, one has to take some operator
ordering and then to come to its discrete representation. A
question arises about the constraints evolution. Constraints can
be violated by both noncommutativity of the operators and
discretization of the equations of motion. The last question is
analogous to that regarding the energy conservation in a system of
the decretized
 magnetic hydrodynamic equations \cite{popov}.
The special discretization schemes were suggested for the magnetic
hydrodynamic equations conserving the energy \cite{popov}.

Let us demonstrate the solution of both problems. The operators
for the Hamiltonian and the momentum constraints could be
discretized by using the symmetrical ordering of operators:

\begin{widetext}
\bea
 \hat {\mathcal P}_j=S\Biggl(
e^{2\hat\alpha_j}\Biggl(-\frac{\hat\alpha_{j+1}-\hat\alpha_j}{3\Delta
x}\,\hat\alpha_j^\prime+\frac{\hat B_{j+1}-\hat B_j}{\Delta x}
\,\hat B_j^\prime+\frac{2}{3}\frac{\hat
\alpha_{j+1}-\hat\alpha_j}{\Delta x}\, \hat B_j^\prime+\frac{\hat
V_{j+1}-\hat V_j}{\Delta x} \,\hat V_j^\prime+\frac{\hat
B^\prime_{j+1}-\hat B^\prime_j}{3\Delta x}+\frac{\hat
\alpha^\prime_{j+1}-\hat \alpha^\prime_j}{3\Delta
x}\Biggr)\Biggr),~\label{pcond}
\\
 \hat {\mathcal H}_j=S\Biggl(\frac{1}{2}e^{2\hat
\alpha_j}\left(-\hat\alpha_j^{\prime 2}+\hat B_j^{\prime 2}+\hat
V_j^{\prime 2}\right) +e^{2\hat\alpha_j+4 \hat
B_j}\Biggl(\frac{1}{6}\left(
\frac{\hat\alpha_{j+1}-\hat\alpha_j}{\Delta x }\right)^2
+\frac{\hat\alpha_{j+1}-2\hat\alpha_j+\hat\alpha_{j-1}}{3(\Delta
x)^2 }+~~~~~~~~~~~~~~~~~~~~~~~~~~~~~~~\nonumber\\\frac{7}{6}\left(
\frac{\hat B_{j+1}-\hat B_j}{\Delta x }\right)^2+\frac{\hat
B_{j+1}-2\hat B_j+\hat B_{j-1}}{3(\Delta x)^2
}+\frac{4}{3}\frac{(\hat \alpha_{j+1}-\hat \alpha_j )(\hat
B_{j+1}-\hat B_j)}{(\Delta x)^2}+\frac{1}{2}\left( \frac{\hat
V_{j+1}-\hat V_j}{\Delta x
}\right)^2\Biggr)\Biggr).~~\label{hamcond}
\eea
\end{widetext}

 Therein and hereinafter we will use the prime to denote a
derivative over conformal time $\eta$. At this stage more careful
definition of the symmetrization \cite{masl,kar} is needed. For an
arbitrary function $f(x_1, x_2, \cdots x_n)$ of
 $n-$variables, one may define a formal Fourier transform
\bea
 \tilde f(\zeta_1,\zeta_2\dots\zeta_n)=~~
 ~~~~~~~~~~~~~~~~~~~~~~~~~~~~~~~~~~~~~~~~~~~~~~~
 \nonumber\\\frac{1}{(2\pi)^n}\int
f(x_1,x_2,\dots x_n)e^{-i\left(x_1\zeta_1+x_2\zeta_2\dots
x_n\zeta_n \right)}d x_1\dots d x_n.\nonumber
 \eea A symmetrized
function of noncommutative operators $\hat A_1, \dots \hat A_n $
is defined as
 \bea   S(f(\hat A_1,\hat A_2\dots \hat
 A_n))=~~~~~~~~~~~~~~~~~~~~~~~~~~~~~~~~~~~~
 \nonumber\\\int
\tilde f(\zeta_1,\zeta_2,\dots \zeta_n)e^{i(\hat A_1\zeta_1+\hat
A_2\zeta_2\dots \hat A_n\zeta_n )}d \zeta_1\dots d \zeta_n.
\nonumber
\eea

Our idea is to use the discretized version of Eqs. (\ref{sv1}),
(\ref{sv2}) as the equations of motion:

\bea
\hat {\mathcal H}_j^\prime=\frac{e^{4 \hat B_{j+1}}\hat{\mathcal
P}_{j+1}+\hat{\mathcal P}_{j+1}e^{4 \hat B_{j+1}} - e^{4 \hat
B_{j}}\hat{\mathcal P}_{j}-\hat{\mathcal P}_{j}e^{4 \hat B_{j}}
}{2 \Delta x },~\label{eqHd}
\\
\hat {\mathcal P}_j^\prime=\frac{\hat{\mathcal
H}_{j+1}-\hat{\mathcal H}_{j}}{3\Delta
x}.~~~~~~~~~~~~~~~~~~~~~~~~~~~~~~~~~~~~~~~~~~~~~\label{eqPd}
\eea

Using the formula for differentiation of a symmetrized function
\cite{kar}:
\bea
 \frac{d}{d\eta} S(f(\hat A_1(\eta),\hat A_2(\eta)\dots \hat
A_n(\eta)))~~~~~~~~~~~~~~~~~~~~~~~~~~\nonumber\\=S\left(\sum_j^n
\frac{d \hat A_j}{d \eta}\,{\partial_j} f(\hat A_1(\eta),\hat
A_2(\eta)\dots \hat A_n(\eta))\right),~~ \label{differ}
\eea (here $\partial_j f(x_1,x_2\dots x_n)$ denotes a partial
derivative of a function $f$ over the $j-$argument) allows
calculating the time derivatives\
 in the left hand side of Eqs.
(\ref{eqHd},\ref{eqPd})
   and rewriting
Eqs. (\ref{eqHd},\ref{eqPd}) in the form of

\begin{widetext}
\bea
  S\Biggl( e^{2\hat \alpha_j}\Biggl(\hat B_j^\prime \hat
B_j^{\prime\prime}+\hat V_j^\prime \hat V_j^{\prime\prime}- \hat
\alpha_j^\prime \hat\alpha_j^{\prime\prime}+ \frac{1}{3(\Delta
x)^2}e^{4\hat B_j}\biggl(7 \hat B_j^2+7\hat B_{j+1}^2+2 \hat
B_{j-1}+3\hat V_j^2
 -2\hat B_j(2+7\hat
B_{j+1} -4\hat\alpha_j+4\hat\alpha_{j+1})
\nonumber\\
+2\hat B_{j+1}(1-4\hat \alpha_j+4\hat\alpha_{j+1})-6\hat V_j\hat
V_{j+1} +3\hat V_{j+1}^2-4\hat \alpha_j+\hat \alpha_j^2 +2\hat
\alpha_{j+1}-2\hat \alpha_j\hat \alpha_{j+1}+\hat \alpha_{j+1}^2
+2\hat \alpha_{j-1}\biggr)\biggl(2 \hat B_j^\prime+\hat
\alpha_j^\prime \biggr)\nonumber\\
 +\hat \alpha_j^\prime\biggl(\hat B_j^{\prime 2}+\hat
V_j^{\prime 2}-\hat \alpha_j^{\prime 2}\biggr)+ \frac{1}{3(\Delta
x)^2} e^{4 \hat B_j}\biggl((-2+7\hat B_j-7\hat B_{j+1}+4\hat
\alpha_j-4\hat \alpha_{j+1})\hat B_j^\prime +(1-7\hat B_j+7 \hat
B_{j+1}-4\hat \alpha_j
\nonumber\\
+4\hat \alpha_{j+1})\hat B_{j+1}^\prime +\hat B_{j-1}^\prime+
3\hat V_j\hat V_j^\prime-3\hat V_{j+1}\hat V_j^\prime-3\hat
V_j\hat V_{j+1}^\prime+3\hat V_{j+1}\hat V_{j+1}^\prime- 2\hat
\alpha_j^\prime+4\hat B_j\hat \alpha_j^\prime-4\hat B_{j+1}\hat
\alpha_j^\prime+\hat \alpha_j\hat \alpha_j^\prime-\hat
\alpha_{j+1}\hat \alpha_j^\prime
\nonumber\\
+ \hat \alpha_{j+1}^\prime-4\hat B_j\hat\alpha_{j+1}^\prime+4\hat
B_{j+1}\hat\alpha_{j+1}^\prime-\hat\alpha_j\hat\alpha_{j+1}^\prime+\hat\alpha_{j+1}
\hat\alpha_{j+1}^\prime +\hat\alpha_{j-1}^\prime\biggr)
 \Biggr)
\Biggr)\nonumber\\=\frac{e^{4 \hat B_{j+1}}\hat{\mathcal
P}_{j+1}+\hat{\mathcal P}_{j+1}e^{4 \hat B_{j+1}} - e^{4 \hat
B_{j}}\hat{\mathcal P}_{j}-\hat{\mathcal P}_{j}e^{4 \hat B_{j}}
}{2 \Delta x },~~~\label{discrh}\\
  S\Biggl(e^{2\hat\alpha_j}\biggl(-3\hat B_j^{\prime2}-3\hat
V_j^{\prime2}+2\hat B_{j+1}^{\prime}\hat \alpha_j^\prime-\hat
\alpha_j^{\prime
2}+2\hat\alpha_j\hat\alpha_j^{\prime2}-2\hat\alpha_{j+1}\hat\alpha_j^{\prime2}
 +3\hat V_j^\prime(\hat V_{j+1}^\prime+2(-\hat V_j+\hat
V_{j+1})\hat\alpha_j^\prime)
+\hat\alpha_j^\prime\hat\alpha_{j+1}^\prime
\nonumber\\
+\hat B_j^\prime(3\hat B_{j+1}^\prime +(-4-6\hat B_j+6\hat
B_{j+1}-4\hat\alpha_j
 +4\hat\alpha_{j+1})\hat\alpha_j^\prime+2\hat\alpha_{j+1}^\prime)+
(-1-3\hat B_j+3B_{j+1})\hat B_j^{\prime\prime} -2\hat\alpha_j\hat
B_j^{\prime\prime} +2\hat\alpha_{j+1}\hat B_j^{\prime\prime}
\nonumber\\
+\hat B_{j+1}^{\prime\prime} +3(-\hat V_j+V_{j+1})\hat
V_j^{\prime\prime}-\hat
\alpha_j^{\prime\prime}+(\hat\alpha_j-\hat\alpha_{j+1})\hat\alpha_j^{\prime\prime}
+\hat\alpha_{j+1}^{\prime\prime} \biggr)\Biggr)={\hat{\mathcal
H}_{j+1}-\hat{\mathcal H}_{j}},~~~\label{discrp}
\eea
\end{widetext}
where $\hat{\mathcal H}_{j}$ and $\hat{\mathcal P}_{j}$ are given
by (\ref{pcond}), (\ref{hamcond}). As the third equation, the
discretized version of Eq. (\ref{10}) can be taken

\begin{eqnarray}
  \hat V^{\prime\prime}_j+S\Biggl(2 \hat
V_j^\prime\hat\alpha_j^\prime-e^{4\hat B}\biggl(\frac{\hat
V_{j+1}-2\hat V_j+\hat V_{j-1}}{(\Delta x
)^2}~~~~~~~~~~~~\nonumber\\
+2\frac{(\hat V_{j+1}-\hat
V_j)(\hat\alpha_{j+1}-\hat\alpha_j)}{(\Delta x)^2
}~~~~~~~~~~~~~~\nonumber\\+4\frac{(\hat V_{j+1}-\hat V_j)(\hat
B_{j+1}-\hat B_j)}{(\Delta x)^2 }\biggr)\Biggr)=0.~~\label{last}
\end{eqnarray}

It is evident that the constraints remain zero during evolution
because $\hat{\mathcal H}_j$, $\hat{\mathcal P_j}$ equal to zero
initially at $\alpha_0\approx -\infty$ in accordance with the
initial conditions and remain zero for this initial conditions
according to Eqs. (\ref{eqHd}), (\ref{eqPd}). These equations are
equivalent to Eqs. (\ref{discrh}), (\ref{discrp}). One has note
that Eqs. (\ref{discrh}), (\ref{discrp}) can be used even  without
right hand side. If another initial conditions are chosen, the
constraints are nonzero and evolve in accordance with Eqs.
(\ref{eqHd}), (\ref{eqPd}).

 Thus,
we have three operator equations
(\ref{discrh}),(\ref{discrp}),(\ref{last}) (or, equivalently, Eqs.
(\ref{discrp}),(\ref{last}) without the right-hand sides), which
are equivalent completely to the classical equations (\ref{11}),
(\ref{12}), (\ref{10}) for commuting quantities in the limit of
$\Delta x \rightarrow 0$ .

The equations should be solved with the following initial
conditions
\[
 \hat V_j(0)=\frac{i}{\Delta x}\frac{\partial}{\partial
\pi_{Vj}},~~~~\hat V_j^\prime(0)=e^{-2\alpha_0}\pi_{Vj},~~
\]
\bea
 \hat
B_j(0)=B_0-\frac{1}{\pi_B}\biggl(\frac{1}{3}\sqrt{\pi_{Vj}^2+\pi_B^2}
~~~~~~~~~~~~~~~~~~\nonumber\\
+\sum_{k=1}^j
S\biggl(\pi_{V\,k-1}\,\frac{i}{\Delta x}\biggl(\frac{\ptl}{\ptl
\pi_{V\,k} }-\frac{\ptl}{\ptl \pi_{V\, k-1}
}\biggr)\biggr)\biggr),
\nonumber\\
 \hat B_j^\prime(0)=e^{-2\alpha_0}\pi_B,~~~~~~~ \hat
\alpha_j(0)=\alpha_0,~~~~~~~~~~~~~~~\nonumber\\\hat
\alpha^\prime_j=e^{-2\alpha_0}\sqrt{\pi_{Vj}^2+\pi_B^2},~~~~
~~~~~~~~~~~~~~~~~~~~~~~~~~ \label{in2}
\eea
where $j\in{1,N}$ and it is implied that $\pi_{V\,0}=\pi_{V\,N}$.

Eqs. (\ref{in2}) are a discrete analog to the initial conditions
(\ref{rlz3}), but the infinite integral for $\hat B(0)$ is
replaced by finite sum, because we use a box with the periodic
boundary conditions.
 The initial conditions (\ref{in2}) have to provide zero
values of the constraints (\ref{pcond}), (\ref{hamcond}) at the
initial moment of time, at $\alpha_0\rightarrow-\infty$. That is
really so for all $j\ne N$, but the value of the momentum
constraint turns out to be equal $\hat {\mathcal P}_N(0)=\Omega$
for $j=N$, where the operator $\hat \Omega$ takes the form
\be
\hat \Omega=\sum_{j=1}^N S\left(i\frac{\ptl}{\ptl \pi_{V\,j} }(
\pi_{V\,j-1}-\pi_{V\,j})\right). \label{omega}
\ee
The operator $\hat \Omega$ is symmetrical over all $\hat
V_j=i\frac{\ptl}{\ptl \pi_{V\,j}}$ and $\pi_{V\,j}$. The equality
$\hat \Omega=0$ cannot be satisfied as the operator equation. That
gives the condition for a state vector of the Hilbert space where
quasi-Heisenberg operators should act:
\be
\Omega|\Psi>=0. \label{omeg0}
\ee
The situation is analogous to that occurring in the light cone
quantization of a closed string \cite{brink}, where the residual
condition from the momentum constraint remains after the excluding
of the superfluous degrees of freedom.  In our case it is possible
to do not impose the condition (\ref{omeg0}) but  to take a large
number of the field oscillators $N$ to diminish a relative error.
This corresponds to the limit $N\rightarrow \infty$ of an infinite
system with existence of some minimal length $\Delta x$.

In the vicinity of small scale factors, the time derivatives are
much larger than the spatial ones so that the operator equations
take a simple form

\bea
  \hat\alpha_j^{\prime\prime}+\hat\alpha_j^{\prime
 2}+
\hat V_j^{\prime
 2}+\hat B_j^{\prime
 2}
 =0,~~~~~~~~~~~~\label{11a}\\
\hat B_j^{\prime\prime}+2 \hat
B_j^\prime\hat\alpha_j^\prime=0,~~~~~~~~~~~~~~~~~~~~~~\label{12a}
\\\hat V_j^{\prime\prime}+2
\hat V_j^\prime\hat\alpha_j^\prime=0,~~~~~~~~~~~~~~~~~~~
\label{10a}
\eea

with the constraint $-\hat\alpha_j^{\prime 2}+\hat B_j^{\prime
2}+\hat V_j^{\prime 2}=0$. All quantities, i.e. $\hat
\alpha_j^{\prime},\hat B_j^{\prime}, \hat V_j^{\prime}$ in these
equations
 commute with each
other. The solution of Eqs. (\ref{11a}),(\ref{11a}),(\ref{12a})
with the initial conditions (\ref{in2}) is given as
\bea
\hat V_j(\eta)= V_j(0)~~~~~~~~~~~~~~~~~~~~~~~~~ ~~~
~~~~~~~~~~~~~~~~~~~~\nonumber\\+\frac{
\pi_{Vj}}{2\sqrt{\pi_B^2+\pi_{Vj}^2}}\ln\left(1+2e^{-2 \alpha_0
}\sqrt{\pi_B^2+\pi_{Vj}^2}\,\eta\right),\nonumber
\eea
\begin{eqnarray*}
 \hat B_j(\eta)=\hat
B_j(0)~~~~~~~~~~~~~~~~~~~~~~~~~~~~~~~~~~~~~~~~~~~~
\\+\frac{
\pi_{B}}{2\sqrt{\pi_B^2+\pi_{Vj}^2}}\ln\left(1+2e^{-2 \alpha_0
}\sqrt{\pi_B^2+\pi_{Vj}^2}\,\eta\right),\\
\hat \alpha_j(\eta)=\alpha_0~~~~~~~~~~~~~~~~~~~~~~~~~
~~~~~~~~~~~~~~~~~~~~~~~~\\+\frac{1}{2}\ln\left(1+2e^{-2 \alpha_0
}\sqrt{\pi_B^2+\pi_{Vj}^2}\,\eta\right).
\end{eqnarray*}

Although the evolution in the vicinity of small $\eta$, where
$\alpha\approx\alpha_0\approx - \infty$ is relatively simple, the
evolution governed by the general equations (\ref{discrh}),
(\ref{discrp}), (\ref{last}), when the fields begin to oscillate
is very complicated and needs the numerical investigation that is
beyond the scopes of this article.

\section{Discussion and Conclusion}

The choice of a considered system state needs an additional
analysis. This state cannot be  identified with ``initial" one.
The reason is that only the state $C(\pi_{V1},\dots...,\pi_{VN})$
describing all evolution of system allows calculating the mean
values of the quasi-Heisenberg operators at any moment of time.
This state is not related to the notion of ``vacuum" since there
is no field oscillators in the limit of $\alpha\rightarrow
-\infty$.

The solution of the Wheeler-DeWitt equation for the Gowdy model
was investigated in Ref. \cite{mizn}. Although, another quantities
were introduced in Ref. \cite{mizn}, instead of logarithm of scale
factor $\alpha$ and $B$-field, namely
\bea
T=B+\alpha,\nonumber\\
\lambda=6(B-\alpha),
\eea
the asymptotic of the Wheeler-DeWitt equation in the vicinity of
$T\rightarrow -\infty$ contains only the momentums by analogy with
Eq. (\ref{witt}) and admits the solutions of a plane wave type.
Asymptotic of the Wheeler-DeWitt equation in the vicinity of
$T\rightarrow \infty$ is of an oscillator type \cite{mizn}.
 Quasiclassical treatment of the Wheeler-DeWitt equation with regard
 to an evolution of universe allows interpreting as
a scattering problem \cite{mizn}. This means that a packet of
plane waves at $T\rightarrow -\infty$ evolves to a number of
gravitons at $T\rightarrow \infty$. It should be noted that there
exists no the state without gravitons at $T\rightarrow \infty$.

The later work \cite{berger} regarding the Gowdy model
quantization seems
 a step back in some sense, because it considers the graviton creation from vacuum in
a style of Refs. \cite{Par69,SexUrb69,ZelSta71,frol}. As was
suggested in Ref. \cite{berger}, there are no gravitons at some
time $T_0$ when the field oscillators exist already and the
gravitons appear later from the vacuum during evolution.

In the presented evolutionary picture, a state describing an
evolution of universe exists in the form of the ``plane waves"
packet. During evolution, this will result in an appearance of
vacuum and some gravitons over it, i.e. one may expect that the
correlators $<\hat V(\eta,x)\hat V(\eta,x^\prime)>$ will be
analogous to the correlators of QFT corresponding to some
gravitons over background.

There is no a wave packet, which would give the correlators
corresponding to a pure vacuum of universe in future, i.e. an
appearance of matter is inevitable in this model. However, the
matter is not created from a vacuum, because universe is not empty
always. It should be reminded that in our model state of a system
does not evolve, we would be only able to calculate the
correlators of the quasi-Heisenberg operators and do some
conclusions by comparing them with the correlators of conventional
QFT.

As an example of a state,
 one may take
\be
C(\pi_{V1},\dots...,\pi_{VN})=C_N\exp\left(-b\sum_{j=1}^N
\pi_{Vj}^2\right),
\ee
where $C_N$ is the normalization constant, and we do not put the
residual condition (\ref{omeg0}) for simplicity.

 This
state implies that the momentums are random and independent in
spatial points. It is not similar to a vacuum state of QFT, as the
fields (and their momentums) in a QFT vacuum state are highly
correlated in the nearest spatial points.

 As an example of the mean
value calculation, one may take the evolution of a ``Hubble
constant" at small $\eta$:
\bea
  <\frac{1}{\hat a_j}\frac{d \hat a_j}{dt}>=<\frac{1}{\hat
a_j^2}\frac{d \hat a_j}{d\eta}>=<\exp(-\hat \alpha_j)\frac{d \hat
\alpha_j}{d\eta}>\nonumber\\=\frac{b^{1/4}}{2
\eta^{3/2}}\,U({1}/{4},{3}/{4},b\, \pi_B^2),
\eea
where $U(a,b,z)$ is the confluent hypergeometric function.

  For this state, the
universe is expanded uniformly in a mean. It would be interesting
to calculate the evolution of correlators $<\hat V_j(\eta)\hat
V_n(\eta)>$. Initially the field $\hat V_j$ is uncorrelated for
the different $j, n$: $<\hat V_j(\eta)\hat V_n(\eta)>\sim
\delta_{jn}$ but then some correlation should arise.

To summarize, the failure of QFT in a flat spacetime to deal with
such an inherently non-linear theory as gravity and existence of
``problem of time"  insists on an invention of some new
quantization procedures. The quasi-Heisenberg quantization scheme
considered may provide the calculational framework for an
investigation of quantum evolution. The goal of further
investigation may be the vacuum energy problem, more exactly, its
possible zero value in the quantization scheme considered. That
may result from the compensation of zero point fluctuations of
gravitational waves by the quantum fluctuations of scale factor.
Thereby, the fluctuations do not contribute to mean evolution. It
should be noted, that this will be purely quantum effect lacking
in classics \cite{riple}. An additional issue is a calculation of
field correlators in order to determine their correspondence to
correlators of the ordinary QFT in late times.

\addcontentsline{toc}{section}{References}
%%%%%%%%%%%%%%%%%%%%%%%%%%%%%%%%%%%%%%%%%%%%%%%%%%%%%%%%%%%%%%%%%%%

\end{document}